\documentclass[10pt, conference, compsocconf]{IEEEtran}
\usepackage{cite}

\usepackage{enumerate}
\usepackage{enumitem}
\usepackage{graphicx}
\usepackage{subcaption}
\usepackage{cleveref}
\Crefformat{figure}{#2Fig.~#1#3}
\Crefmultiformat{figure}{Figs.~#2#1#3}{ and~#2#1#3}{, #2#1#3}{ and~#2#1#3}
\begin{document}
\title{A Review of Star Schema Benchmark}
\author{\IEEEauthorblockN{Jimi Sanchez}
\IEEEauthorblockA{Department of Computer Science\\
East Carolina University\\
Greenville, NC USA\\
%Email: sanchezji14@students.ecu.edu
}
}

% * <jimi.linuxguy@gmail.com> 2015-12-24T16:34:29.329Z:
%
% ^.
\maketitle
\begin{abstract}
This paper examines the Star Schema Benchmark, an alternative to the flawed TPC-H decision support system and presents reasons why this benchmark should be adopted over the industry standard for decision support systems.
\end{abstract}
\renewcommand\IEEEkeywordsname{Keywords}
\begin{IEEEkeywords}
TPC-H; databases; star schema benchmark; DBMS; profiling;
\end{IEEEkeywords}

\section{INTRODUCTION}
The TPC organization is responsible for defining benchmarking standards. One such standard is TPC-H, an ad hoc decision support benchmark \cite{Nambiar2009PerformancePapers}. TPC-H has been criticized for not adhering to a Ralph Kimball model of data marts, not adhering to Edgar F. Codd’s definition of a 3NF data schema, as well as not allowing freedom in indexing and partitioning \cite{Coronel2009DatabaseManagement,Date2013RelationalProfessionals,Nambiar2006TheTPC-DS}. The Star Schema Benchmark (SSB) was designed to test star schema optimization to address the issues outlined in TPC-H with the goal of measuring performance of database products and to test a new materialization strategy. The SSB is a simple benchmark that consists of four query flights, four dimensions, and a simple roll-up hierarchy \cite{ONeil2009PerformanceBenchmarking}. The SSB is significantly based on the TPC-H benchmark with improvements that implements a traditional pure star-schema and allows column and table compression.
\par
The SSB is designed to measure performance of database products against a traditional data warehouse scheme. It implements the same logical data in a traditional star schema whereas TPC-H models the data in pseudo 3NF schema \cite{RochaAlvaroCorreiaAnaMariaCostanzoSandraReis2015New1}.
\section{Compression}
Typically higher degrees of correlation can be found between columns of transactional data. Highly correlated columns are often able optimized using compression routines and column stores. However due to the data distribution, it is impractical and unrealistic to attempt to use compression on the TPC-H schema as can be seen in \Cref{section:Scaling}. SSB allows the columns in tables to be compressed by whatever means available in the database system used, as long as reported data retrieved by queries has the values specified in the schema definition \cite{ONeil2009PerformanceBenchmarking}. It has been shown that compressing data using column-oriented compression algorithms as well as keeping the data in a compressed format may improve performance significantly by as much as an order of magnitude \cite{2014BigPapers}. Data stored in columns is more compressible than data stored in rows. This is due to the fact that compression algorithms perform better on data with low information entropy. Entropy in data compression refers to the randomness of the data being passed into the compression algorithm. The higher the entropy, the lower the compression ratio. In other words, the more random the data is the harder it becomes to compress it \cite{Aghav2010DatabaseOptimization}.
\par
This observation only immediately affects compression ratio, %Disk space is considered to be cheap and has continued to become cheaper with each passing year. Compression improves performance as well as reducing used disk space. 
less time is spent in input/output activities as data is read from disk into memory if the data is compressed \cite{Abadi2006IntegratingSystems}. Some heavier compression algorithms that optimize for compression ratio may not be as suited in comparison to lighter schemes that forfeit compression ratio for decompression performance. If a query executor can operate directly on compressed data, then decompression can be avoided completely, furthermore, decompression can be avoided then performance can be further improved. One example would be the query executor having the ability to perform the same operation on multiple column values at once \cite{Abadi2006IntegratingSystems}.
\par
It has been shown that the most significant compression differences between row and column stores are the cases where a column has been sorted and contains consecutive repeating of some value in that column \cite{Abadi2006IntegratingSystems}. It is extremely easy to summarize these value repeats in a column-store. It is even easy to operate directly on this summary. In contrast, in a row-store, surrounding data from other attributes significantly complicates the process. For this reason, compression will typically have a more significant impact on query performance when the percentage of columns accessed have some order \cite{Abadi2006IntegratingSystems}.
\begin{figure}
\includegraphics[width=0.5\textwidth]{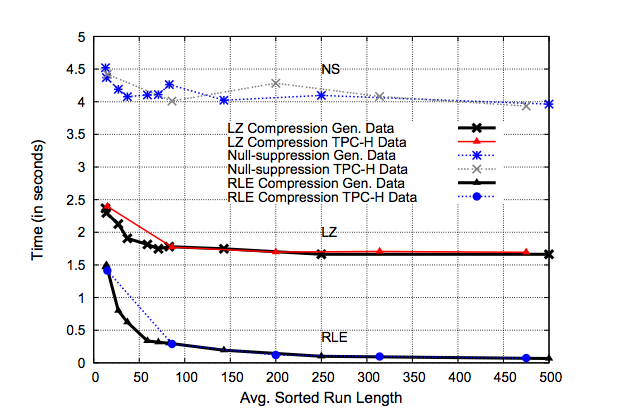}
\caption{Comparison of query performance on data}
\label{fig:Compression_Results}
\end{figure}
\Cref{fig:Compression_Results} shows the comparison of query performance on TPC-H and generated data using RLE, LZ, and null-suppression compression schemes. 
\section{Detail}
Schema modifications were made to the TPC-H schema transform it into the more efficient star schema form. The TPC-H tables \textit{LINEITEM} and \textit{ORDERS} are combined into one sales fact tale named \textit{LINEORDER}. \textit{LINEORDER} is consistent with a denormalized warehouse per the Kimball model \cite{Kimball2013TheModeling}. Kimball claims that a star schema helps to reduce the number of complex and often unnecessary joins \cite{Kimball2013TheModeling}.
\par
The \textit{PARTSUPP} table is dropped since it would belong to a different data mart than the \textit{ORDERS} and \textit{LINEITEM} data, furthermore, it contains temporal data that varies. The comment attributes for \textit{LINEITEMS}, \textit{ORDERS}, and shipping instructions are also dropped as a data warehouse does not store such information in a fact table, they can't be aggregated and take significant storage space.
\par
A dimension table called \textit{DATE} is added to the schema as is in line with a typical data warehouse. However, as this is a commonly reserved word in many DBMS systems, a more relevant name can be used to avoid SQL errors or having to wrap the table name in back-tick identifiers. These table simplifications result in a proper star schema data mart. \textit{LINEORDER} serves as a central fact table. Dimension tables are created for \textit{CUSTOMER}, \textit{PART}, \textit{SUPPLIER}, and \textit{DATE}. A series of tables for \textit{SHIPDATE}, \textit{RECEIPTDATE}, and \textit{RETURNFLAG}, should also be constructed, but would result in a too complicated a schema for our simple star schema benchmark. SSBM concentrates on queries that select from the \textit{LINEORDER} table exactly once. It prohibits the use of self-joins or sub-queries as well as or table queries also involving \textit{LINEORDER}. The classic warehouse query selects from the table with restrictions on the dimension table attributes. SSBM supports queries that appear in TPC-H. SSB consists of one large fact table (\textit{LINEORDER}) and four dimensions tables (\textit{CUSTOMER}, \textit{SUPPLIER}, \textit{PART}, \textit{Date}). It is common practice to combine \textit{LINEITEM} and \textit{ORDER} in TPC-H to get \textit{LINEORDER} in SSB. \textit{LINEORDER} represents one row for each one in \textit{LINEITEM}.
\begin{figure}[!htb]
\includegraphics[width=0.5\textwidth]{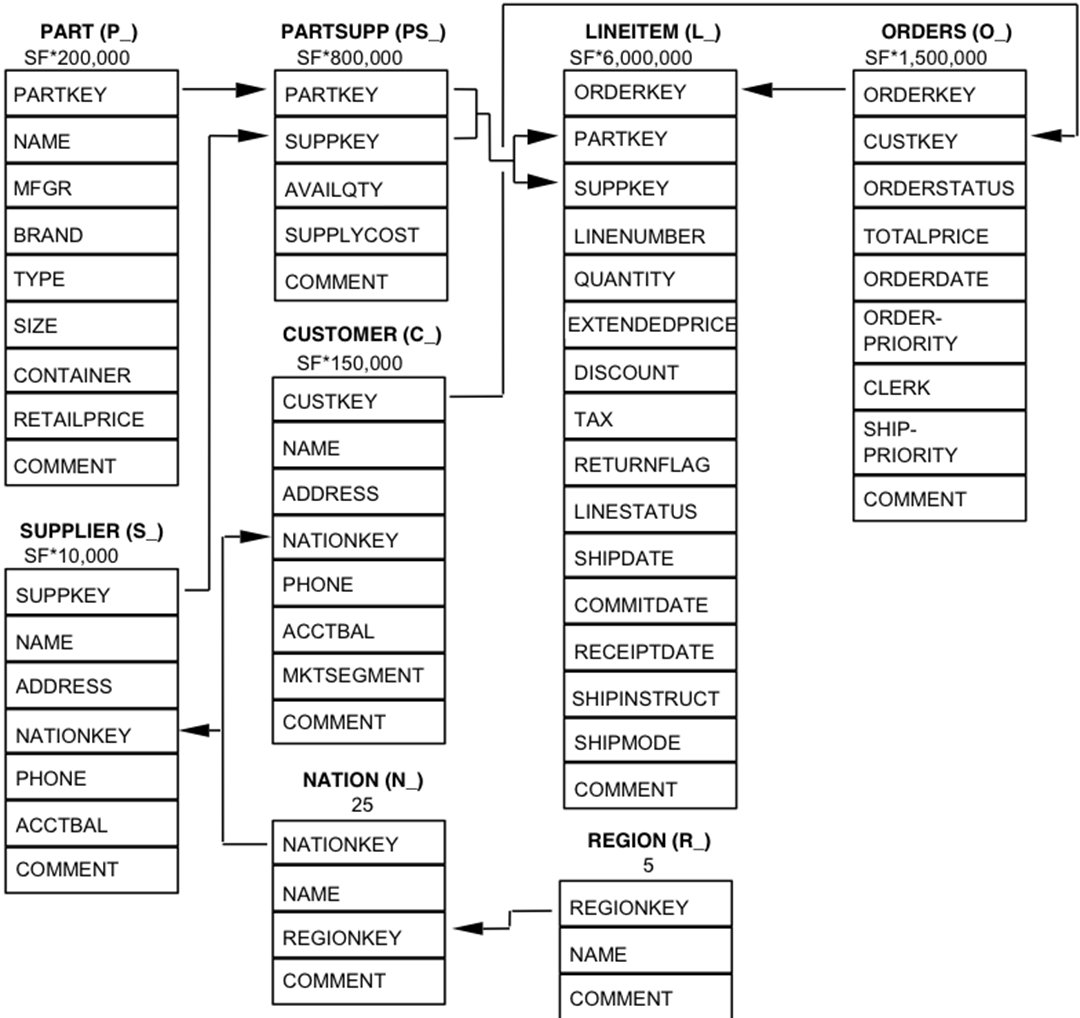}
\caption{TPC-H Schema}
\label{fig:TPCH_Schema}
\end{figure}
\begin{figure}
\includegraphics[width=0.5\textwidth]{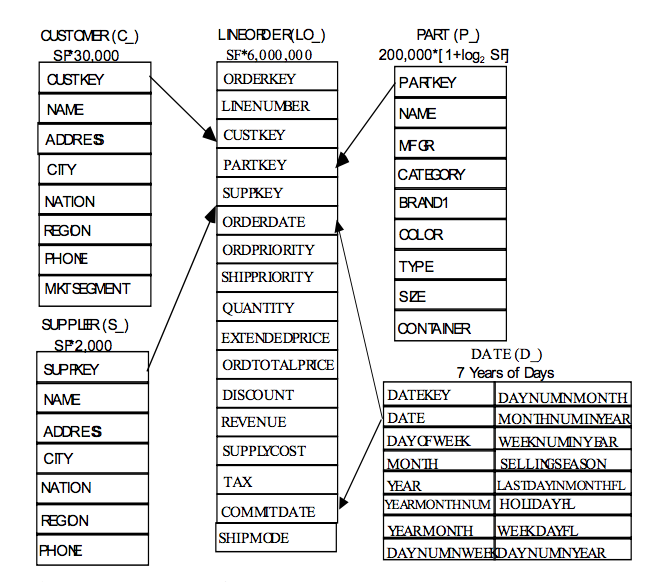}
\caption{Star Schema Benchmark Implementation}
\label{fig:SSB_Schema}
\end{figure}
\subsubsection{PARTSUPP}
As seen in \cref{fig:SSB_Schema} the \textit{PARTSUPP} table was removed to keep in line with the data warehousing principles of Kimball. The problem with this methodology is that the \textit{LINEITEM} and \textit{ORDERS} tables have fine temporal grain. The \textit{PARTSUPP} table has a Periodic Snapshot grain. Periodic snapshot grains are fact tables that summarize many measurement events occurring over a some period of time, such as a day, a week, or a month. Subsequently, transactions that add new rows over time to \textit{LINEORDER} do not modify rows in \textit{PARTSUPP}. One solution would be to treat \textit{PARTSUPP} and \textit{LINEORDER} as separate fact tables, isolating queries separately and not joined together. This is done in all but one of the queries where \textit{PARTSUPP} is in the WHERE clause (Q1, Q11, Q16, Q20) but not in Q9, where \textit{PARTSUPP}, \textit{ORDERS}, and \textit{LINEITEM} all appear \cite{Council2011TPC-HSpecification}. Q9 is intended to find, for each nation and year, the profits for certain parts ordered that year \cite{Council2011TPC-HSpecification}. The problem is that it is beyond the bounds of reason that the \textit{PS\_AVAILQTY} would have remained constant during all these past years. This difference in grain between \textit{PARTSUPP} and \textit{LINEORDER} is what causes the problem. One reason for having the PARTSUPP is to break up what might be a star schema and so that query plans do not appear to be too simple.
\par
Creating a Snapshot on the \textit{PARTSUPP} table seems to be overly complicated as to create a non-trivial join that was designed to complicate the query path and add more load to the system artificially. In the TPC-H benchmark \textit{PS\_AVAILQTY} is never updated, not even during the refresh that inserts new rows into the ORDERS table. In SSB data warehouse, it is more reasonable to leave out the \textit{PARTSUPP} tale and instead create a new column \textit{SUPPLYCOST} for each \textit{LINEORDER} Fact row. A data warehouse contains derived data only, so there is no reason to normalize in order to guarantee one fact in one place.
\par
It is possible, and perhaps likely that subsequent orders for the same part and supplier might repeat this \textit{SUPPLYCOST}. If the last part of some kind were to be deleted its reasonable to believe that it might result in the loss of the price charged. Since SSB is attempting to simplify queries this might be an acceptable solution. In fact, SSB adds the \textit{LO\_PROFIT} column to the \textit{LINEORDER} table which aids in making these calculations simpler and execution time quicker.
\subsubsection{LINEORDERS}
The \textit{LINEITEM} and \textit{ORDERS} table are combined into one sales fact table that is named \textit{LINEORDER}. This is a standard denormalization aligned with the data warehousing per Kimball \cite{Kimball2013TheModeling}. This combination of tables into one sales fact table reduces the need for many complex joins spread across the most common queries. All columns in the \textit{ORDERS} and \textit{LINEITEMS} tables that make us wait to insert a Fact row after an order is placed on \textit{ORDERDATE} is dropped. An example is not wanting to wait until we know when the order is shipped, when it is received, and whether it is returned before we can query the existence of an order. 
\subsection{NATION and REGION}
The \textit{NATION} and \textit{REGION} tables are denormalized into the \textit{CUSTOMER} and \textit{SUPPLIER} tables and a \textit{CITY} column is added. This simplifies the schema considerably, both for writing queries and computing queries as the two largest tables of TPC-H are pre-joined. Queries do not have to perform the join and users writing queries against the schema do not have to express the join in their queries. \textit{NATION} and \textit{REGION} might be appropriate in an OLTP system to enforce integrity, but not in a data warehouse system where the data is cleaned before being loaded and dimension tables are not so limited in space use as the fact table. \textit{NATION} and \textit{REGION} are added to the \textit{ADRESSS} columns. 
\section{Queries}
The reasons for the departure from the TPC-H query format are multiple. The most common reason is an attempt to provide the Functional Coverage and Selectivity Coverage features explained in Set Query Benchmark \cite{ONeil1991TheBenchmark., ONeil2009PerformanceBenchmarking}. The benchmark queries are chosen as much as possible to span the tasks performed by an important set of Star Schema queries, so that prospective users can derive a performance rating from the weighted subset they expect to use in practice. It is difficult to provide true functional coverage with a small number of queries, but SSB at least tries to provide queries that have up to four dimensional restrictions. Selectivity coverage is the idea that the total number of fact table rows retrieved will be determined by the selectivity of restrictions on the dimensions. SSB introduces  variety into the selectivity by varying the queries results sets. The goal of SSB is to provide functional coverage as well as selectivity coverage. Some model queries are based on the TPC-H query set, but need to be modified so vary the selectivity. These are the only queries that will not have an equal counterpart in TPC-H.
\subsubsection{Query Flights}
Compared to the TPC-H's 22 queries, SSB contains four query flights that each consist of three to four queries with varying selectivity. Each flight consists of a sequence of queries that someone working with data warehouse systems would ask such as a dill down \cite{ONeil2009PerformanceBenchmarking}.
\subsubsection{Caching}
One other issue arises in running the Star Schema Benchmark queries, and that is the caching effect that reduces the number of disk accesses necessary when query Q2 follows query Q1, because of overlap of data accessed between Q1 and Q2. SSB attempts to minimize overlap, however in situations where this cannot be done, SSB will take whatever steps are needed to reduce caching effects of one query on another \cite{ONeil2009PerformanceBenchmarking}.
\section{Data Distribution}
Similar to TPC-H all of the data in SSB is uniformly distributed. Selectivity hierarchies are introduced in all dimension tables similar to the manufacturer/brand hierarchy in TPC-H \cite{ONeil2009PerformanceBenchmarking}. This uniform data distribution is aided by a tool called SSB-DBGEN. SSB-DBGEN is tool similar to TPC-H DBGEN that makes populating the database for benchmark runs simple and allow for quick transitions between transaction tests. However, SSB-DBGEN is not easy to adapt to different data distributions as its metadata and actual data generation implementations are not separated \cite{Rabl2013VariationsPerformance}.
\section{Scaling}\label{section:Scaling}
Both TPC-H and SSB generate data at different scales using the scaling factor. The scale factor determines the amount of information initially loaded into the benchmark tables. Scale factor increases the size of the database during the testing process. As the scale factor increases, the number of rows added to the tables increase. Data is generated proportionally to scale factor. The scale factor impacts the number of generated lines. Only for the \textit{PARTS}, data is not scaled linearly but logarithmically \cite{ONeil2009PerformanceBenchmarking}.
\section{Experiments}
\subsection{Out of the Box Configuration}
Utilizing an Amazon Web Services (AWS) Elastic Compute Cloud (EC2) m1.small instance running CentOS 6.5, we installed MySQL 5.6.12, PostgreSQL 9.4, and SQLite 3.9.2. We created two databases on each DBMS, one for testing the TPC-H schema and one for the SSB schema. We then used the DBGEN and SSB-DBGEN tools respectively to populate the newly created databases with data using a scale factor of 1. TPC-H and SSB both use a base scale factor which can be used to scale the size of the benchmark. The sizes of each of the tables are defined relative to this scale factor. Using QGEN and SSB-QGEN, we generated randomized TPC-H and SSB queries. Next we mapped 10 TPC-H queries to the closest matching SSB query flight and query. Some TPC-H have direct mapping such as TPCHQ3 and TPCH6. The remaining queries that did not have direct TPC-H equivalents were matched based on the business logic of the query, as well as the complexity of the query. No indices were added to the schemes except those created by the DBMS for primary keys. Our query pair mapping of TPC-H to SSB can be seen in \cref{fig:queryMapping}.
\begin{figure}[!htb]
  \centering
%  \begin{table}
  \begin{tabular}{| l | c | r |}
  \hline
  Query & TPC-H & SSB \\
    \hline
    1 & Q6 & Q1.1 \\
    2 & Q6 & Q1.2 \\
    3 & Q6 & Q1.3 \\
    4 & Q3 & Q5.1 \\
    5 & Q3 & Q5.2 \\
    6 & Q3 & Q5.3 \\
    7 & Q2 & Q12.1 \\
    8 & Q2 & Q12.2 \\
    9 & Q5 & Q13.1 \\
    10 & Q5 & Q13.2 \\
  \hline
  \end{tabular}
 % \end{table}
  \caption{TPC-H to SSB Mapping}
  \label{fig:queryMapping}
\end{figure}

\begin{figure}[!htb]
\centering
\includegraphics[width=0.5\textwidth]{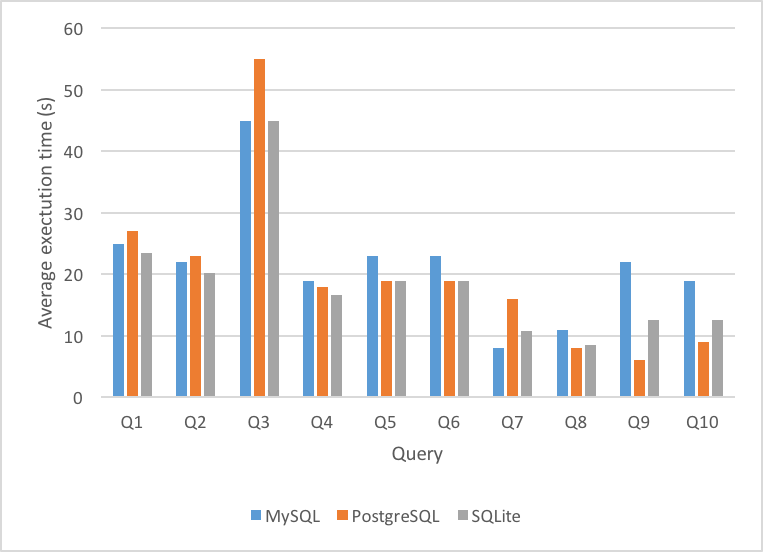}%
\caption{TPC-H average execution time (Out of the box configuration)}%
\label{fig:tpc-out-of-box-sf1}%
\end{figure}

\begin{figure}
\centering
\includegraphics[width=0.5\textwidth]{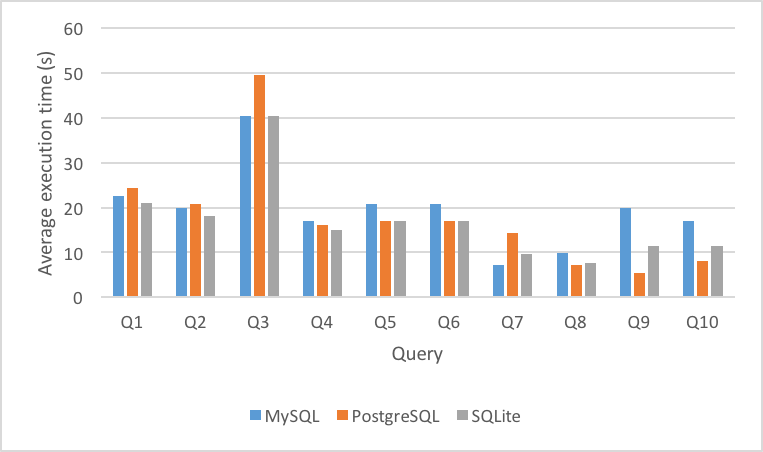}
\caption{SSB average execution time (Out of the box configuration)}
\label{fig:ssb-out-of-box-sf1}%
\end{figure}

\Cref{fig:tpc-out-of-box-sf1} shows the average execution time of each of the first ten queries of TPC-H executed on each DBMS.
\Cref{fig:ssb-out-of-box-sf1} shows the average execution time of each of the first ten queries of SSB executed on each DBMS. It is clear from this graph that there is a significant decrease in average execution time across the board from TPC-H. In fact, there is an increase in execution time in all queries across all DBMS.

\par
As can be seen from the data, even though the average execution time decreased significantly across the board from TPC-H to SSB, the ratio between each DBMS maintained. 
\subsection{Indexing}
Next we analyzed the WHERE clauses and EXPLAIN output from each of the 10 queries in both TPC-H and SSB and created indices on those columns that would benefit from an index (all clauses specified in the WHERE clause as well as some composite keys). 

\begin{figure}[!h]
\centering
\includegraphics[width=0.5\textwidth]{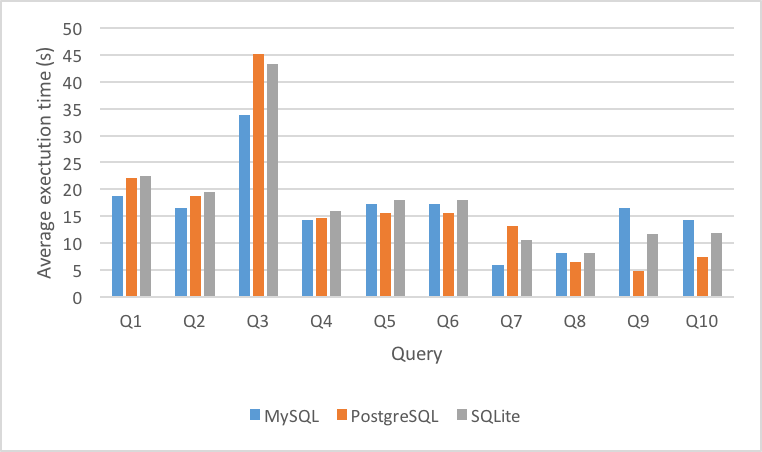}
\caption{TPC-H average execution time (With indices applied)}
\label{fig:tpc-index-sf1}
\end{figure}

\begin{figure}[!h]
\includegraphics[width=0.5\textwidth]{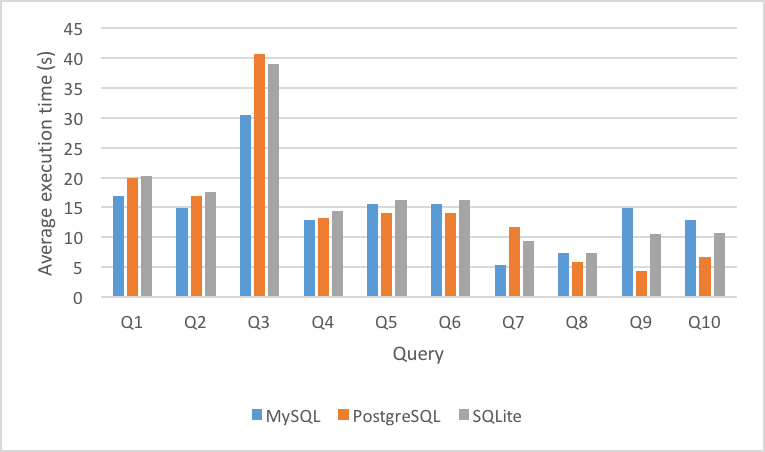}
\caption{SSB average execution time (With indices applied)}
\label{fig:ssb-index-sf1}
\end{figure}

\Cref{fig:tpc-index-sf1} and \Cref{fig:ssb-index-sf1} show the average execution time per DBMS across the systems with indices added for queries Q1-Q10 in TPC-H and SSB. As can be seen in this data, adding indices changed the average execution time significantly, but also changed the results of the DBMS performance now. MySQL is the clear winner for speed alone after utilizing indices, whereas SQLite appeared to be the faster solution without index optimization.

\section{Conclusion}
SSB is a variation of the TPC-H benchmark modified so that it represents a Kimball style database design. SSB is a popular benchmark for decision support systems and is a valid tool for evaluating the performance of DBMS executing star schema queries. Today's systems rely heavily on sophisticated cost-based query optimizers to generate the most efficient query execution plans. SSB evaluates the optimizer's capability to generate optimal execution plans under all circumstances. 
\par
The optimizer needs to be aware of the performance implications of operating directly on compressed data in its cost models. Further, cost models that only take into account I/O costs will likely perform poorly in the context of column-oriented systems since CPU cost is often the dominant factor.
\par
Our experiment has successfully shown that SSB is a far better benchmark, it offers a much simpler schema and query execution set. SSB adheres to the Ralph Kimball definition of a data warehouse. TPC-H penalizes systems who are unable to optimize for correlated sub-queries, limited indexing, and partitioning. SSB helps to level the playing field so that IT decision makers and software architects can make informed choices when deciding which DBMS solution(s) to use.
\bibliography{Mendeley}
\bibliographystyle{IEEEtran}
\end{document}